\documentclass[published]{nst}

\usepackage{subfigure,dcolumn}
\usepackage[T2A,T1]{fontenc}
\usepackage[russian,english]{babel}

\usepackage{listings}
\begin{document}

\title{Design and simulation of the High-Energy Proton Beam Telescope}
\thanks{Supported by the National Natural Science Foundation of China (No.11875274) and (No.U1232202)}

\author{Lan-Kun Li}
\affiliation{Institute of High Energy Physics, Chinese Academy of Sciences, Beijing 100049, China}
\affiliation{Spallation Neutron Source Science Center, Dongguan 523000, China}

\author{Ze Gao}
\affiliation{Zhengzhou Railway Vocational and Technical College, Zhengzhou 450018, China}

\author{Ying-Hao Yu}
\affiliation{Institute of High Energy Physics, Chinese Academy of Sciences, Beijing 100049, China}
\affiliation{University of Chinese Academy of Sciences, Beijing 100049, China}

\author{Liang-Cheng-Long Jin}
\affiliation{Institute of High Energy Physics, Chinese Academy of Sciences, Beijing 100049, China}

\author{Ming-Yi Dong}
\email[Corresponding author, ]{Ming-Yi Dong, Institute of High Energy Physics, No. 19B Yuquan Road, Shijingshan District, Beijing, dongmy@ihep.ac.cn }
\affiliation{Institute of High Energy Physics, Chinese Academy of Sciences, Beijing 100049, China}
\affiliation{University of Chinese Academy of Sciences, Beijing 100049, China}

\author{\\Ren-Hong Liu}
\affiliation{Institute of High Energy Physics, Chinese Academy of Sciences, Beijing 100049, China}
\affiliation{Spallation Neutron Source Science Center, Dongguan 523000, China}

\author{Hong-Yu Zhang}
\affiliation{Institute of High Energy Physics, Chinese Academy of Sciences, Beijing 100049, China}

\author{Chang Xu}
\affiliation{Institute of High Energy Physics, Chinese Academy of Sciences, Beijing 100049, China}
\affiliation{University of Chinese Academy of Sciences, Beijing 100049, China}

\author{Han-Tao Jing}
\affiliation{Institute of High Energy Physics, Chinese Academy of Sciences, Beijing 100049, China}
\affiliation{Spallation Neutron Source Science Center, Dongguan 523000, China}

\author{Yu-Hang Guo}
\affiliation{Institute of High Energy Physics, Chinese Academy of Sciences, Beijing 100049, China}
\affiliation{Spallation Neutron Source Science Center, Dongguan 523000, China}

\author{Qun Ou-Yang}
\affiliation{Institute of High Energy Physics, Chinese Academy of Sciences, Beijing 100049, China}
\affiliation{University of Chinese Academy of Sciences, Beijing 100049, China}

\begin{abstract}
A high-resolution beam telescope is essential for the precise characterization of silicon pixel sensors. As part of the CSNS-II upgrade project, a High-Energy Proton Beam Telescope (HEPTel) based on monolithic active pixel sensors (MAPS) has been designed for the forthcoming High-Energy Proton Experimental Station (HPES), which will provide 0.8–1.6 GeV single-particle proton beams. HEPTel consists of six ultra-thin telescope modules, with a material budget per module of about 0.061\% $X_0$. Simulated with a 1.6 GeV proton beam, the telescope is expected to achieve a resolution of about 1.83 $\mu$m. Additionally, a dedicated readout electronics system and a Data Acquisition (DAQ) system have been designed for HEPTel, based on which a preliminary test system was established for beam tests. The beam test results with 1.3 GeV electrons demonstrated a single-module resolution of about 5.77 $\mu$m, an overall telescope resolution of about 2.70 $\mu$m, and a detection efficiency above 99.5\%. These results validate the HEPTel design and confirm its capability for forthcoming proton-beam experiments at HPES.

\end{abstract}


\maketitle

\section{Introduction}
In the development of new silicon pixel sensors and detector prototypes, a collimated monoenergetic charged-particle test beam is essential for prototype validation. Such facilities deliver high-energy beams that enable precise evaluations of detector characteristics, including spatial resolution and detection efficiency. Various test beam facilities have been established, such as the DESY \uppercase\expandafter{\romannumeral2} test beam facility~\cite{DESY,DESY1} providing 1–6 GeV electrons or positrons, the CERN-SPS accelerator with 20–400 GeV beams~\cite{CERN-SPS,CERN-SPS1}, and the Fermilab Test Beam Facility~\cite{FermiLab} delivering 120 GeV protons and secondary particles. During test beam experiments, beam telescopes serve as the core detectors, offering precise reference tracks for the device under test (DUT) and thus requiring a spatial resolution better than that of DUTs. Different beam telescopes are employed at these test beam facilities, such as the EUDET-type beam telescopes at CERN and DESY~\cite{EUDET1,EUDET2}, the LYCORIS beam telescope equipped with a 1T solenoid magnet at DESY~\cite{Lycoris,Lycoris1}, and the CMS pixel telescope at the Fermilab Test Beam Facility~\cite{CMS2}.

The China Spallation Neutron Source (CSNS)~\cite{CSNS1,CSNS2} upgrade project CSNS-II~\cite{CSNS3} proposes a High-Energy Proton Experimental Station (HPES), which will provide a single-particle proton beam with energies ranging from 0.8 to 1.6 GeV. The HPES will be equipped with multiple detector systems for the development of detectors intended for next-generation collider and cosmic-ray experiments, such as the Circular Electron-Positron Collider (CEPC)~\cite{CEPC1,CEPC2,CEPC3}, the Super Tau-Charm Facility (STCF)~\cite{STCF1,STCF2}, and the High Energy cosmic-Radiation Detection (HERD) facility~\cite{HERD1,HERD2}, among others.

To meet the experimental requirements of HPES, a High-Energy Proton Beam Telescope (HEPTel) has been designed, consisting of six silicon pixel telescope modules. It features an ultra-low material budget to mitigate the effect of multiple Coulomb scattering and to provide a telescope resolution better than 10 $\mu$m for 1.6 GeV protons. The system architecture and simulated performance of HEPTel are presented in Section 2. Section 3 describes the design of the system components in detail, while Section 4 reports the preliminary test results.

\section{System Design and Simulation}
\subsection{High-Energy Proton Experimental Station}

\begin{figure}[htp]
    \centering
    \includegraphics[width=0.45\textwidth]{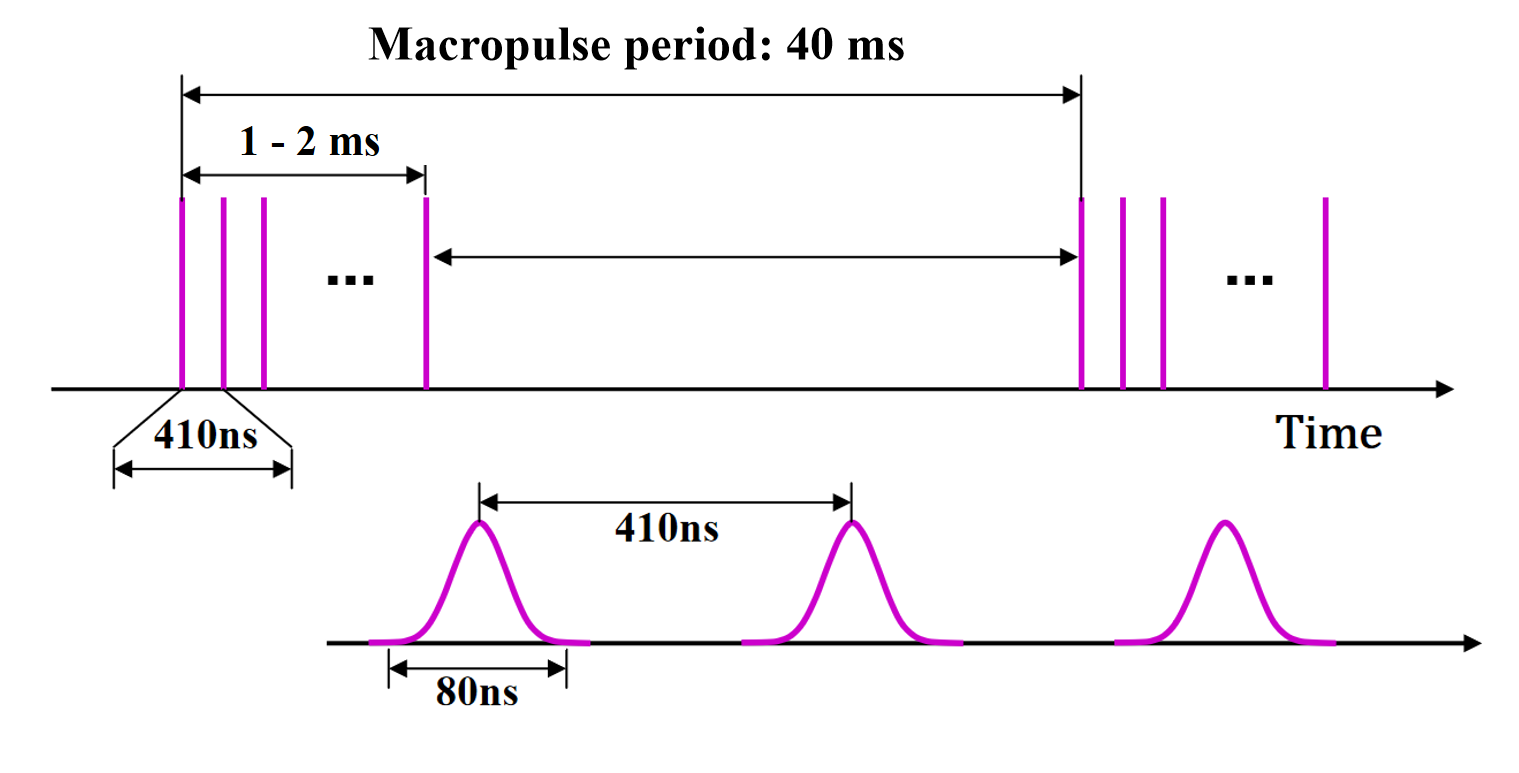} 
    \caption{Schematic of the proton bunch structure and the micropulses time structure at HPES, showing macropulses of 1–2~ms duration repeating every 40~ms.}
    \label{fig:1.1}
    
\end{figure}

At HPES, protons will be extracted from the Rapid Cycling Synchrotron at extremely low intensity using a scattering slow extraction method~\cite{HPES_Beam}. A thin scattering foil is placed at the extraction point, where incident particles collide with the foil and are subsequently extracted at low intensity. The foil rotates around the beam axis at a frequency of 25~Hz, producing macropulses of about 1–2~ms duration that repeat every 40~ms, as shown in Fig.~\ref{fig:1.1}. During each macropulse, the circulating protons revolve in the synchrotron with a revolution period of about 410~ns, which sets the minimum spacing between successive micropulses. This time structure implies that the instantaneous event rate within each macropulse is much higher than the average event rate.

Downstream collimators with adjustable apertures are then used to generate a single-particle proton beam. By tuning the collimator apertures, the typical time interval between protons within a macropulse can be adjusted in the range of a few microseconds to several tens of microseconds. The expected beam spot size is about 20 $\times$ 20 mm$^2$. Furthermore, a degrader allows the beam energy to be adjusted continuously across the 0.8–1.6~GeV range for the application of detector energy calibrations.

The HPES will be equipped with five detector systems to support different DUT tests, including a beam telescope, a proton energy detector based on low-gain avalanche diodes (LGADs), a trigger system, a proton profile detector, and a proton intensity monitor. Among them, the trigger system plays a key role in the operation of HEPTel. It consists of three scintillating-fiber detectors placed upstream and downstream of the telescope, together with a Trigger Logic Unit (TLU) that performs coincidence of trigger signals and generates trigger IDs for event-level synchronization and alignment between the detector systems and the DUTs.

In addition, HPES provides a modular Data Acquisition (DAQ) system that enables the independent configuration and parallel readout of multiple detector systems, along with online data checking and real-time monitoring. These beam conditions and facility infrastructure impose key requirements on HEPTel, including a low material budget to mitigate the effect of multiple Coulomb scattering and reliable track-finding efficiency under high event rate conditions.

\subsection{System architecture of HEPTel}
HEPTel mainly consists of six telescope modules, an electronics and DAQ system, and a high-precision experimental platform. In the baseline design, each telescope module employs a MAPS-based silicon pixel sensor MIMOSA-28 (also known as ULTIMATE)~\cite{MIMOSA28,MIMOSA28_1} with a pixel pitch of 20.7 $\mu$m, which was specifically developed for the Heavy Flavor Tracker (HFT) of the STAR experiment at RHIC.
The detailed design of the telescope modules and other components are described in the following section.

\begin{figure}[htp]
    \centering
    \begin{tabular}{cc}
       \includegraphics[width=0.45\textwidth]{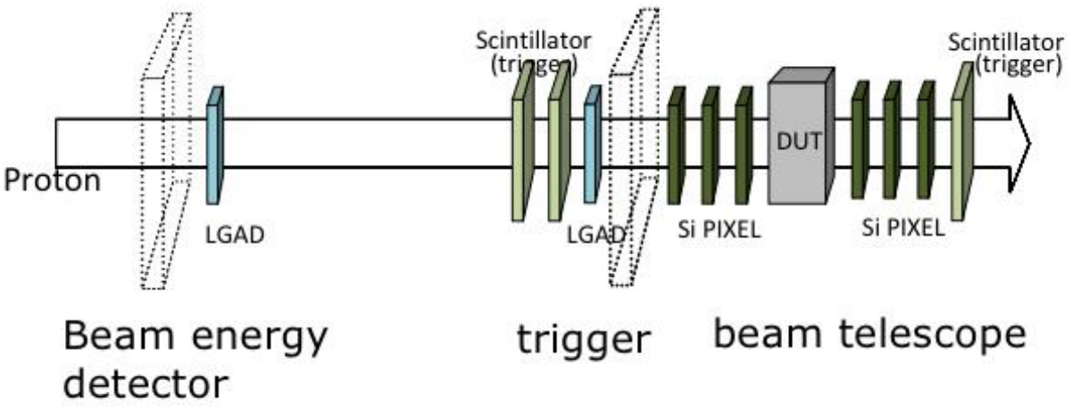} 
    \end{tabular}
    \caption{Schematic layout of the HEPTel setup at HPES, illustrating the relative positions of the beam telescope, trigger detectors, and other detector systems.}
    \label{fig:1.2}
\end{figure}

In a typical setup, the DUT is positioned at the center of the telescope, with three telescope modules placed upstream and three downstream, as shown in Fig.\ref{fig:1.2}. Protons traverse vertically through the six modules and the DUT. During beam tests, residuals between the extrapolated track positions on the DUT plane and the measured DUT hits are used to determine the DUT's spatial resolution and detection efficiency. Since these residuals contain contributions from both the DUT and the telescope, the telescope must provide superior spatial resolution to enable an accurate performance evaluation.

\subsection{Simulation of HEPTel}

To evaluate the performance of HEPTel and guide its design optimization, simulations were performed using Allpix$^2$~\cite{Allpix1,Allpix2}, an open-source software framework for simulating individual silicon pixel chips and more complex detector systems, such as beam telescopes. A six-layer beam telescope and a DUT were simulated with protons. Unless otherwise stated, the DUT and telescope modules were modeled identically, each represented by a silicon pixel sensor with a pitch of 20.7 $\mu$m and a material budget per module of about 0.061\% $X_0$. The spacing between the DUT and its nearest telescope modules, as well as the spacing among the telescope modules, was set to 25 mm. For the simulated data, detector alignment was performed by minimizing track residuals from numerous tracks~\cite{Alignment}, and reference tracks were reconstructed using a Kalman filter with a covariance matrix incorporating multiple Coulomb scattering~\cite{Kalman}.

\begin{figure}[htp]
    \centering
       \includegraphics[width=0.45\textwidth]{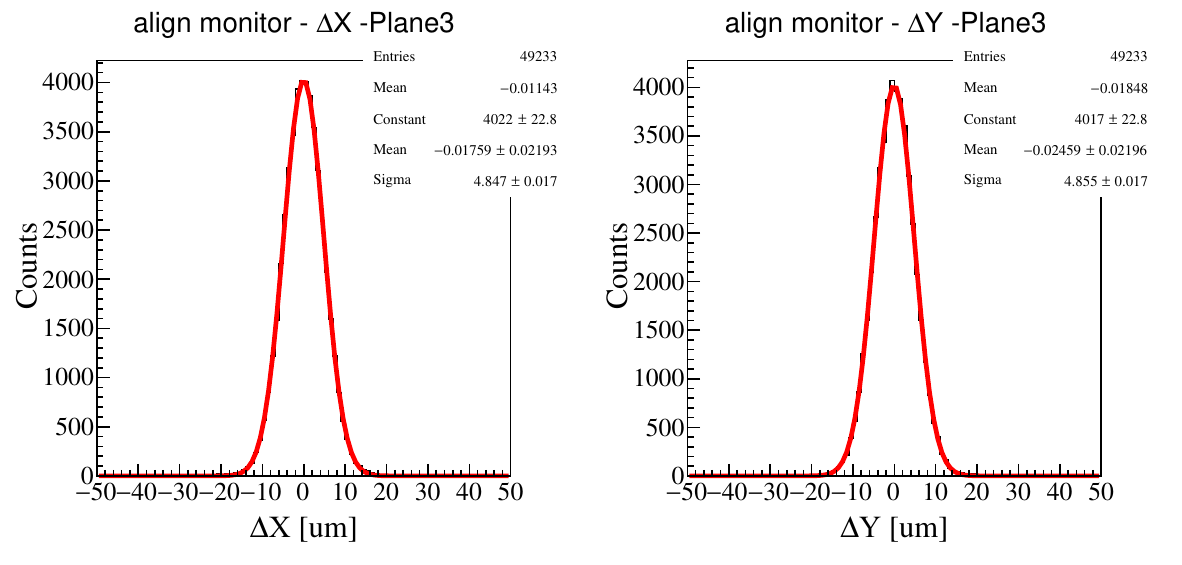} 
    \caption{The simulated residual distributions of the DUT in the X (left) and Y (right) directions using 1.6~GeV protons.}
    \label{fig:1.3}
\end{figure}

In beam tests, the measured residual width of the DUT $\sigma_{meas}$ includes contributions from both the DUT and the telescope~\cite{Resolution}. Assuming identical spatial resolution for all telescope modules and a symmetric arrangement around the DUT, the DUT resolution $\sigma_{\mathrm{DUT}}$ and the telescope resolution $\sigma_{\mathrm{tel}}$ can be calculated using Eq.~\eqref{001}:
\begin{equation}
\label{001}
\begin{aligned}
&\sigma_{\mathrm{DUT}}^2=\frac{N}{N+1}\sigma_{\text {meas }}^2\\
&\sigma_{\text {tel }}^2=\frac{1}{N+1} \sigma_{\text {meas }}^2
\end{aligned}
\end{equation}
where N is the number of telescope modules. Figure~\ref{fig:1.3} shows the simulated residual distribution of the DUT in the X and Y directions. For 1.6 GeV protons, the simulated $\sigma_{meas}$ is about 4.85 $\mu$m, corresponding to a DUT resolution of about 4.48~$\mu$m and a telescope resolution of about 1.83~$\mu$m.

\begin{figure}[htp]
    \centering
       \includegraphics[width=0.45\textwidth]{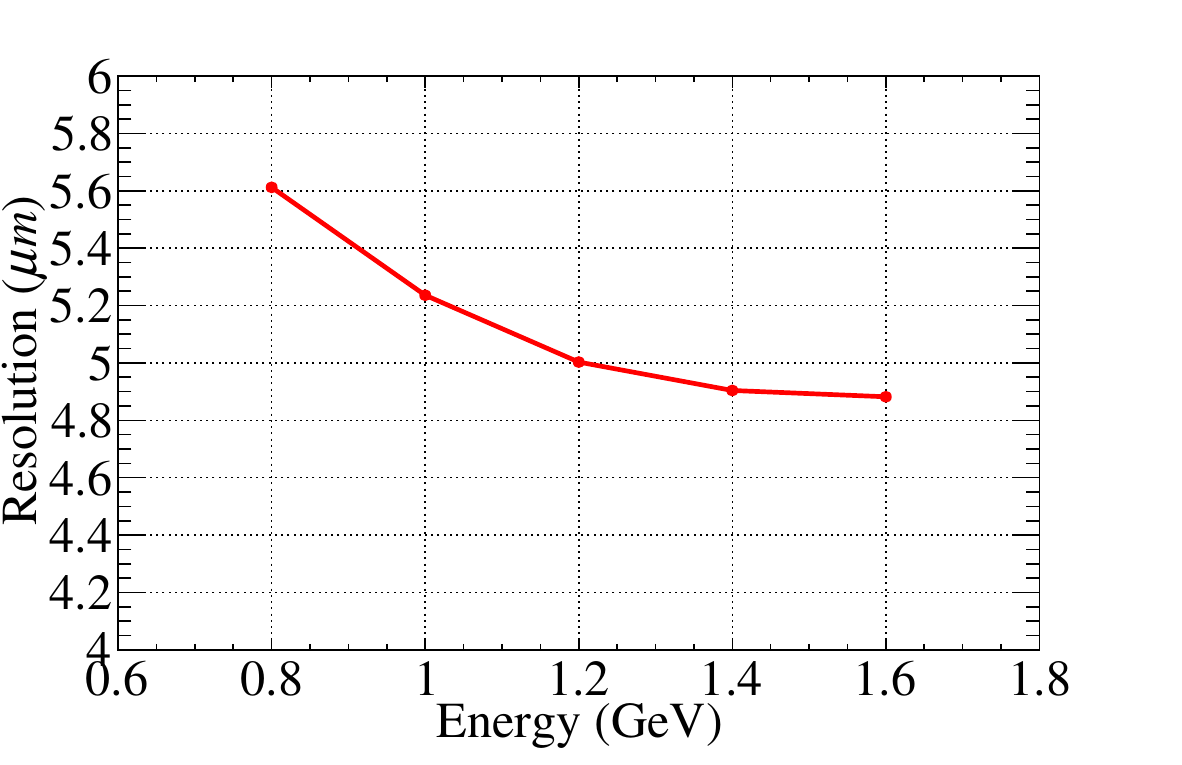} 
    \caption{The simulated residual width as a function of the proton beam energy ranging from 0.8 to 1.6 GeV.}
    \label{fig:1.4}
\end{figure}

The simulated residual width as a function of the beam energy is shown in Fig.~\ref{fig:1.4}. Due to the energy limitation at HPES, multiple Coulomb scattering has a non-negligible influence~\cite{Highland}. Increasing the beam energy can significantly enhance the telescope’s measurement precision. Therefore, beam tests using HEPTel should be performed with 1.6 GeV protons to mitigate the effect of multiple Coulomb scattering.

\begin{figure}[htp]
    \centering
    \begin{tabular}{cc}
         \includegraphics[width=0.45\textwidth]{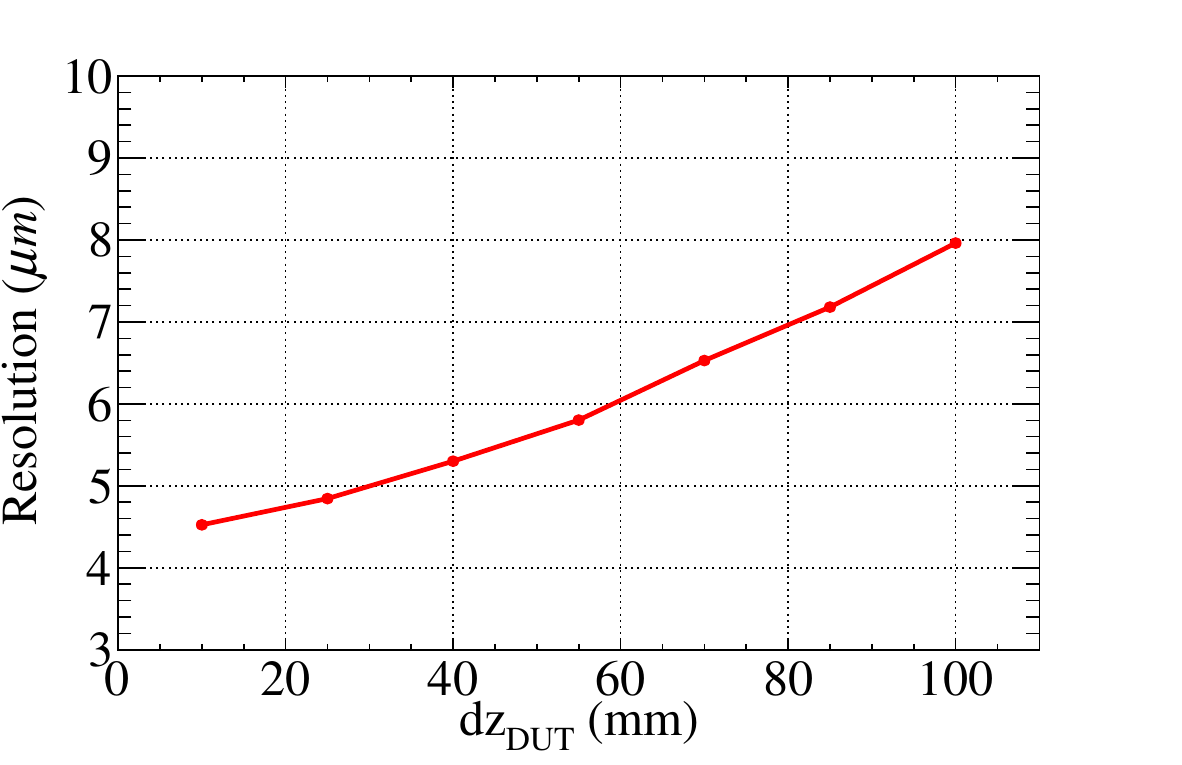}  \\ (a) \\
       \includegraphics[width=0.45\textwidth]{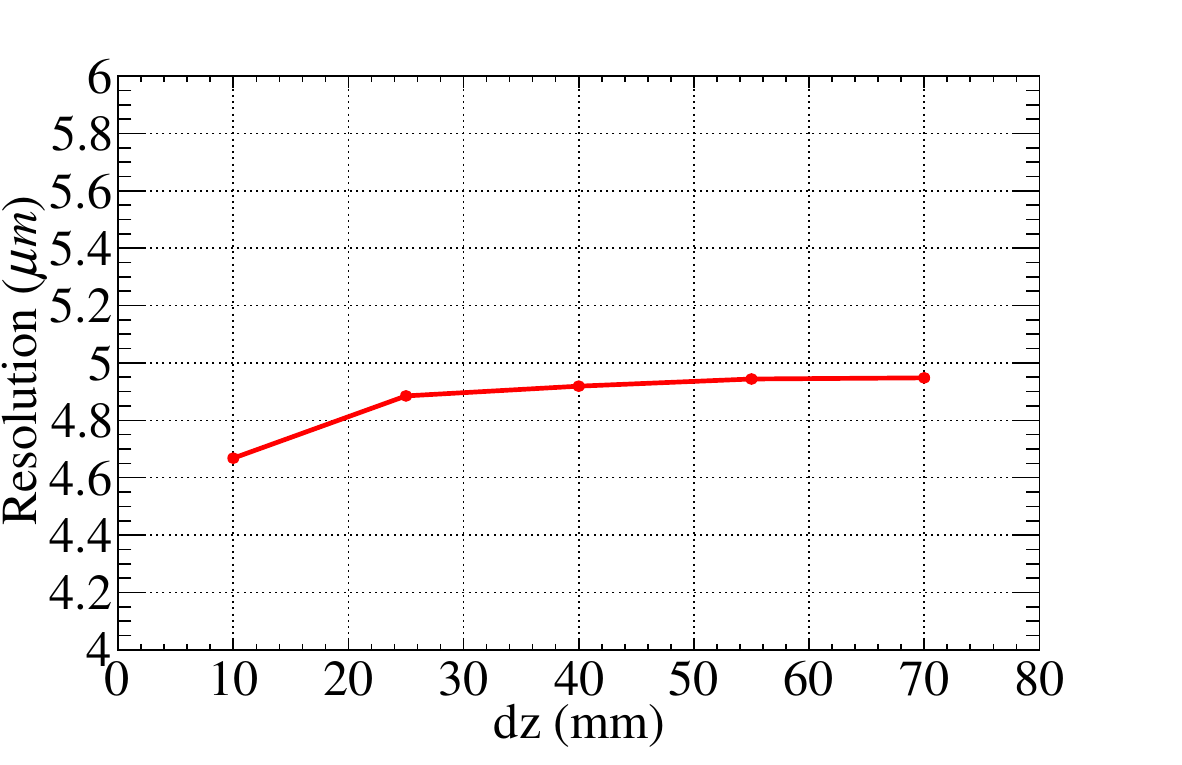}  \\ (b) 
    \end{tabular}
    \caption{(a) The simulated residual width of the DUT versus the distance between the DUT and its nearest telescope planes. (b) The simulated residual width of the DUT versus the distance between telescope planes.}
    \label{fig:1.5}
\end{figure}

The distance between telescope planes can be adjusted along the beam axis to accommodate different DUT sizes. Even with a fixed scattering angle, the detector layout still influences the effect of multiple Coulomb scattering~\cite{Effect}. Figures~\ref{fig:1.5}(a) and~\ref{fig:1.5}(b) show the simulated residual width versus the distance from the DUT to the nearest telescope planes and versus the spacing among telescope planes, respectively. 
The measured accuracy is highly sensitive to the DUT–telescope spacing but only weakly affected by the spacing among telescope planes.
Within 10–100~mm, reducing the DUT–telescope gap significantly mitigates multiple scattering and improves spatial resolution. These simulation results provide essential guidance for the mechanical design of the system. All telescope modules will be mounted on precision sliding rails along the beam axis, enabling flexible adjustment of their positions to accommodate DUTs of various sizes. This configuration allows the telescope modules to be positioned as close as possible to the DUT, thereby minimizing multiple scattering and improving measurement precision.

\subsection{Efficiency of track-finding}
As part of the baseline design of HEPTel, the telescope modules are equipped with MIMOSA-28 sensors operated in a rolling-shutter mode with a frame readout time of 185.6~$\mu$s. At HPES, the proton spacing is adjustable; during HEPTel beam tests, the average event rate is typically set to about 1–5~kHz, corresponding to a mean spacing of a few tens to a few of microseconds. Consequently, each telescope frame often contains multiple tracks. To maintain reliable performance under such conditions, efficient track-finding algorithms are required, and two different track-finding strategies have been therefore investigated.

In the first method, seed tracks are initially formed by pairing hits from Layer-1 and Layer-3. Each seed is extrapolated to Layer-2 and subsequently to the downstream layers (Layer-4 to Layer-6) in sequence. When hit points can be matched, the closest one is added to the seed track, followed by a linear track fit. This iterative process continues until all six layers contribute, at which point the six hits are identified as a candidate track. In the second method, independent linear fits are performed for the upstream and downstream three layers, respectively. The two seed tracks are extrapolated to the DUT plane. If their intersection points can be matched, the six hits are identified as a candidate track. For both methods, once a candidate track is found, the remaining hits are further processed to search for additional tracks.

\begin{figure}[htp]
    \centering
   
         \includegraphics[width=0.45\textwidth]{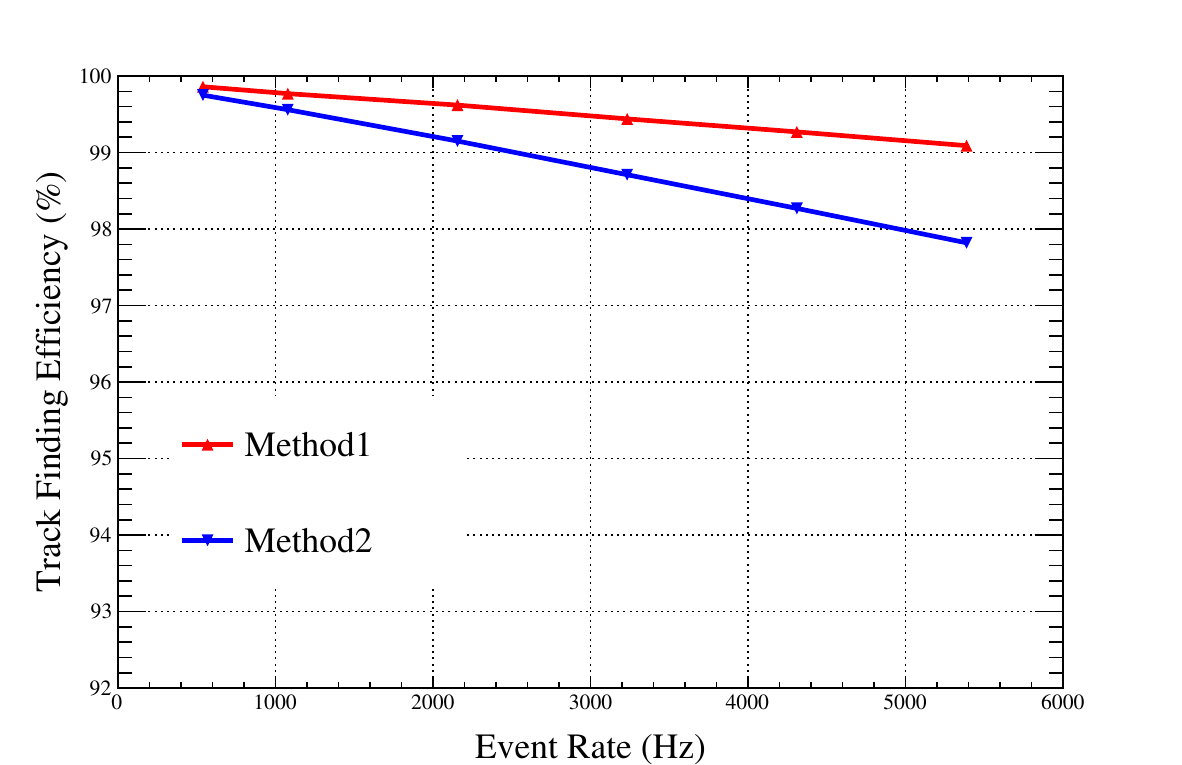}  
     
    \caption{Track-finding efficiency of the two algorithms as a function of the average event rate, ranging from 1 kHz to 5 kHz.}
    \label{fig:1.6}
\end{figure}

Simulations were performed to evaluate the track-finding efficiency of the two algorithms under different event rates. The beam  spot size was set to 10 mm $\times$ 10 mm. The efficiency of each algorithm was determined by comparing the candidate tracks found by the algorithms with the simulated truth tracks. The simulation results show that the first method achieves significantly higher track-finding efficiency than the second method, as illustrated in Fig.~\ref{fig:1.6}. When the average event rate is below 5~kHz, the efficiency of the first method exceeds 99$\%$, ensuring that the track-finding procedure does not introduce significant additional uncertainties, even under high-occupancy conditions. Together with the resolution simulations presented above, these results validate the MIMOSA-28-based baseline design of HEPTel and demonstrate that it meets the performance requirements of HPES. In future upgrades, HEPTel is planned to adopt next-generation silicon pixel sensors with improved time resolution.

\section{HEPTel components}
\subsection{Telescope Module}
Given the HPES beam energy limitation, each telescope module is designed to minimize its material budget in order to reduce multiple Coulomb scattering and improve the overall telescope resolution. Each module consists of a MIMOSA-28 chip thinned to 50 $\mu$m, an auxiliary PCB board, and an aluminum shielding box. The MIMOSA-28 chip is fabricated using 350~nm CMOS technology and contains 928 rows and 960 columns of pixels with a pitch of 20.7~$\mu$m. This corresponds to a sensitive area of about 3.8~cm$^2$, which closely matches the beam spot size at HPES.

\begin{figure}[htp]
    \centering
    \begin{tabular}{cc}
         \includegraphics[width=0.2\textwidth]{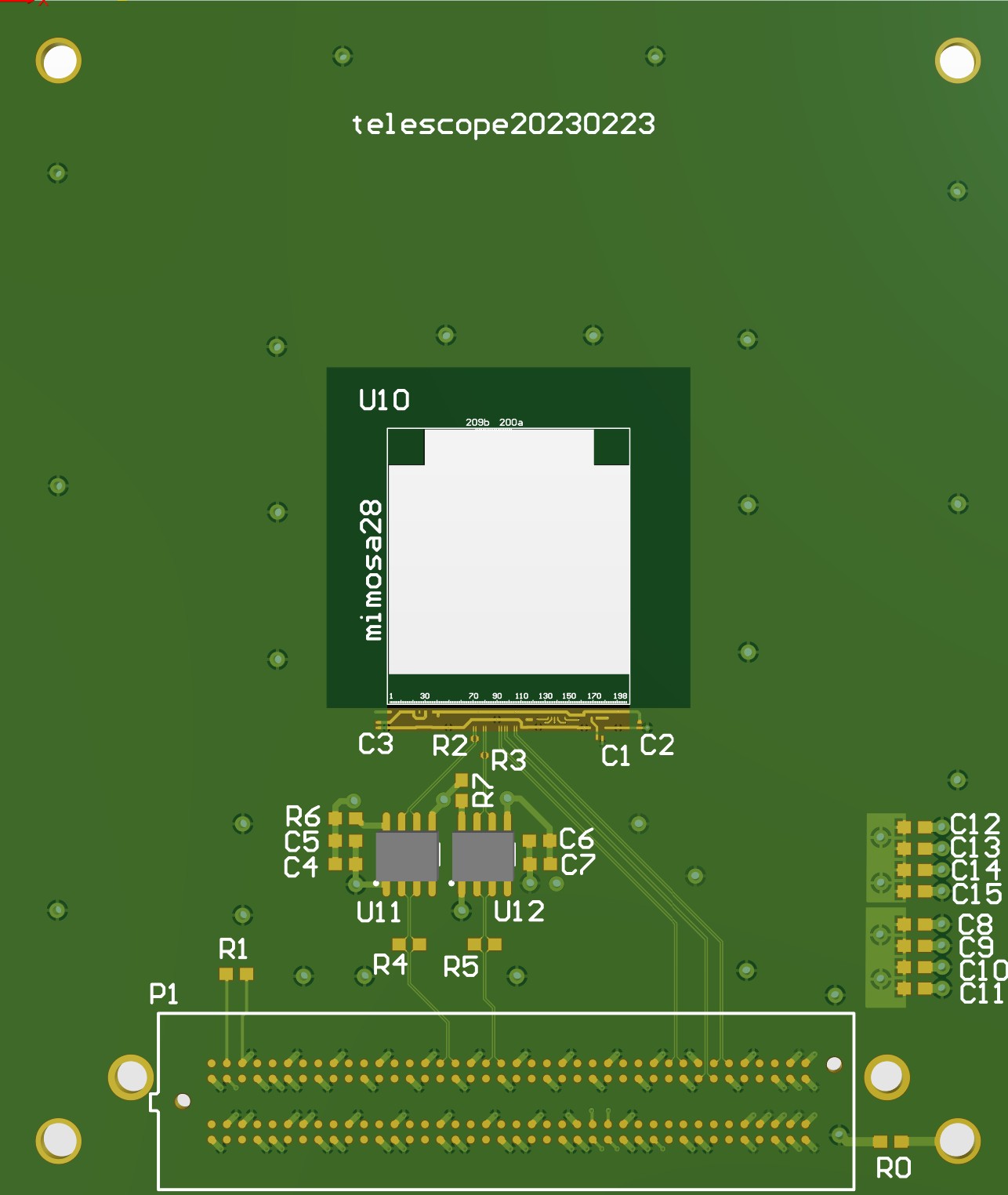}  &
       \includegraphics[width=0.2\textwidth]{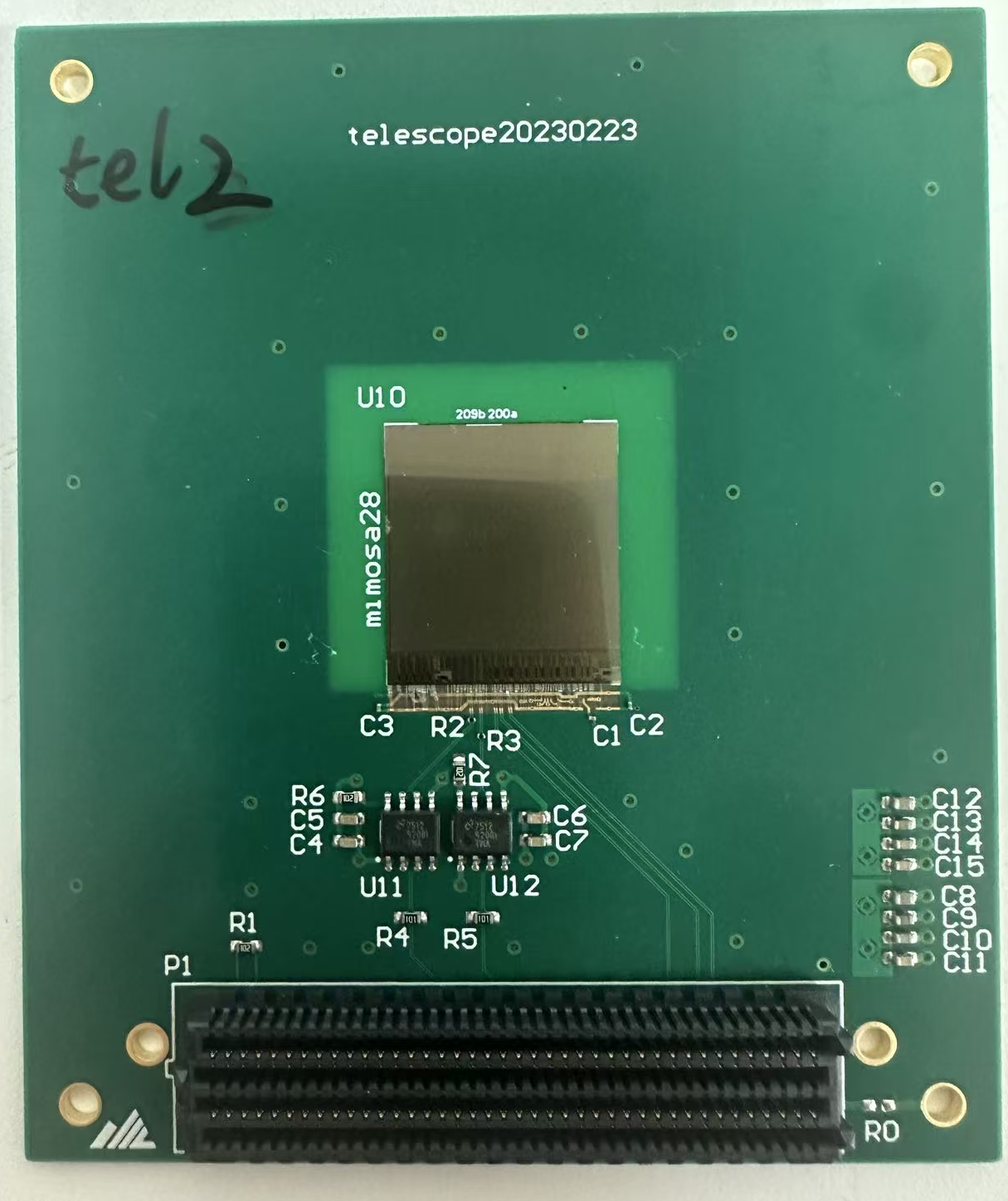}  \\ (a) & (b) 
    \end{tabular}

    \caption{Layout (a) and photograph (b) of the auxiliary PCB for the telescope module.}
    \label{fig:2.1}
\end{figure}

An auxiliary PCB board was specifically designed to interface the telescope module with the electronics system. Only passive components are used to minimize the impact of high-energy protons. To minimize the material budget in the sensitive area, the PCB beneath the sensor is completely hollowed out, leaving only two 3~mm × 3~mm pads at the upper corners and a 2.5~mm strip along the wire-bonding side for mechanical support and electrical connection, as shown in Fig.~\ref{fig:2.1}. The electrical connections between the sensor and the PCB are implemented via wire bonding, and the auxiliary board is then connected to the readout board through an FPGA Mezzanine Card (FMC) connector.

An aluminum shielding box encloses and protects both the MIMOSA-28 chip and the auxiliary PCB board. Square cutouts are made on both sides of the shielding box to correspond to the sensor area, and two layers of black polyimide film are used for light shielding. The total material budget of each telescope module is about 0.061\%~$X_0$, including a 50~$\mu$m silicon chip and two 12.5~$\mu$m polyimide films. The overall thickness of the shielding box is 25 mm. Since the MIMOSA-28 chip is enclosed inside the shielding box, water-cooling is used to dissipate the heat generated by the sensor. A copper cooling pipe is embedded into the bottom part of the box, with thermally conductive silicone grease applied to improve heat transfer. The readout board is mounted upside down onto the FMC interface, while the main heat-generating components are placed outside the shielding box, allowing temperature control via air cooling.

A dedicated assembly jig was designed and produced to ensure precise mechanical alignment between the sensor and the auxiliary PCB board. The jig includes an L-shaped alignment bar, a vacuum chuck, and four positioning pins. During assembly, the MIMOSA-28 chip is first aligned against the L-shaped bar and fixed in position using the vacuum chuck. After the chip is aligned and secured, the auxiliary PCB is inverted and positioned with the aid of the positioning pins, and then glued to the sensor using a 25 $\mu$m-thick acrylic adhesive film. This assembly procedure provides reliable mechanical alignment and sufficient stability for beam-test operation while preserving the low material budget of the module.

\subsection{Electronics and DAQ system}
\begin{figure}[htp]
    \centering
       \includegraphics[width=0.45\textwidth]{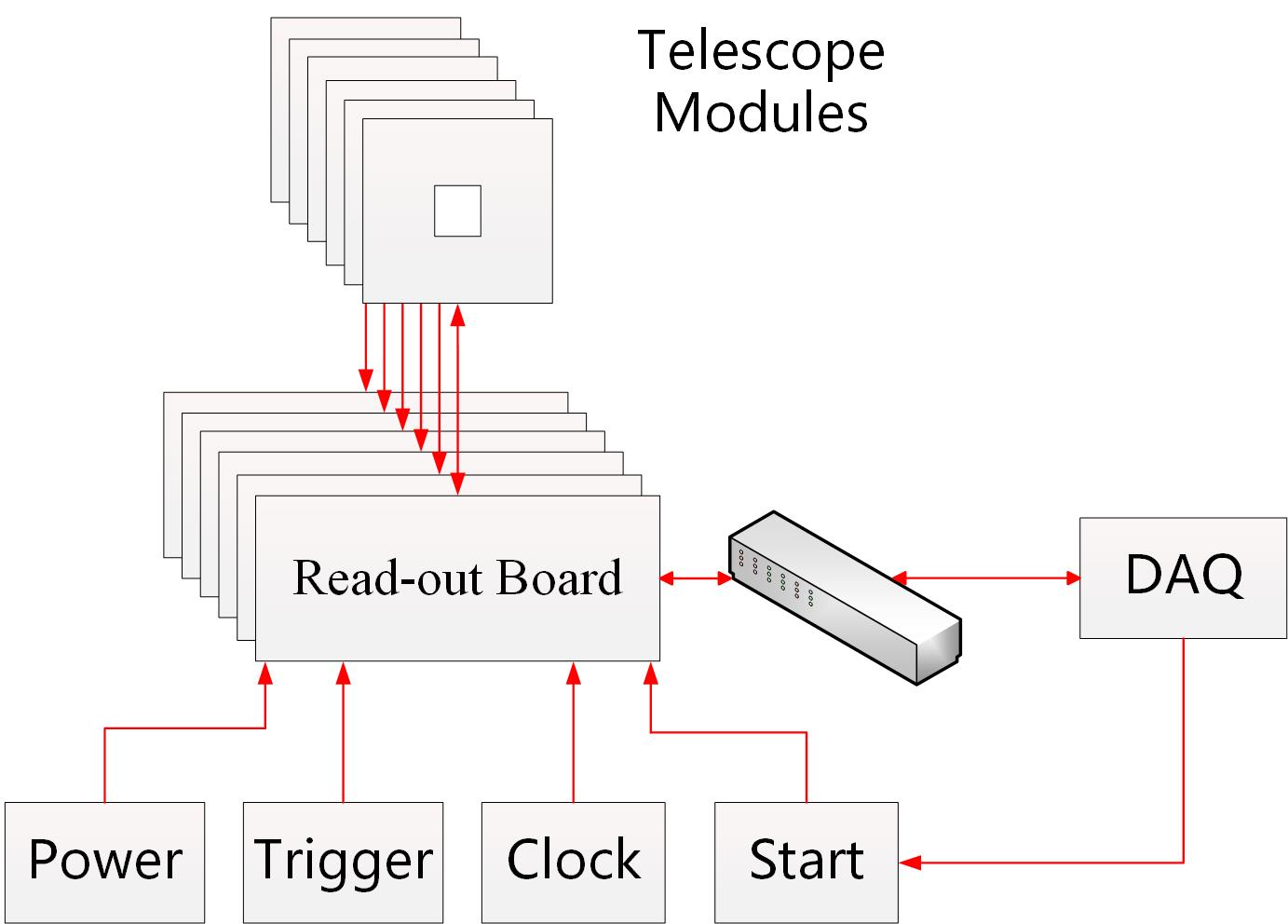} 
    \caption{Schematic diagram of the electronics and DAQ system.}
    \label{fig:2.2}
\end{figure}

The schematic of the electronics system is shown in Fig.~\ref{fig:2.2}. The system adopts a modular architecture, enabling independent control of all six telescope layers. Each module is equipped with a dedicated readout board that interfaces to the auxiliary PCB via an FMC connector, supplying power and clock signals to the sensor and receiving raw data. An HDMI interface on each readout board receives trigger IDs from the TLU for event-level synchronization. Data communication with the DAQ system is implemented through SiTCP-based Gigabit Ethernet, which supports both configuration commands and data transfer. For HEPTel operation, the DAQ system enables independent configuration and parallel readout of the telescope modules, together with online data checking and real-time monitoring to ensure stable operation under high event rate conditions.

A dedicated control board receives the Start/Stop control signals from the DAQ system and distributes them to all six readout boards through a fan-out board, ensuring the synchronous operation of the entire telescope. To maintain clock consistency across all modules, an external clock fan-out board provides a 100 MHz reference clock to each readout board.

\subsection{High-precision experimental platform}
\begin{figure}[htp]
    \centering
       \includegraphics[width=0.45\textwidth]{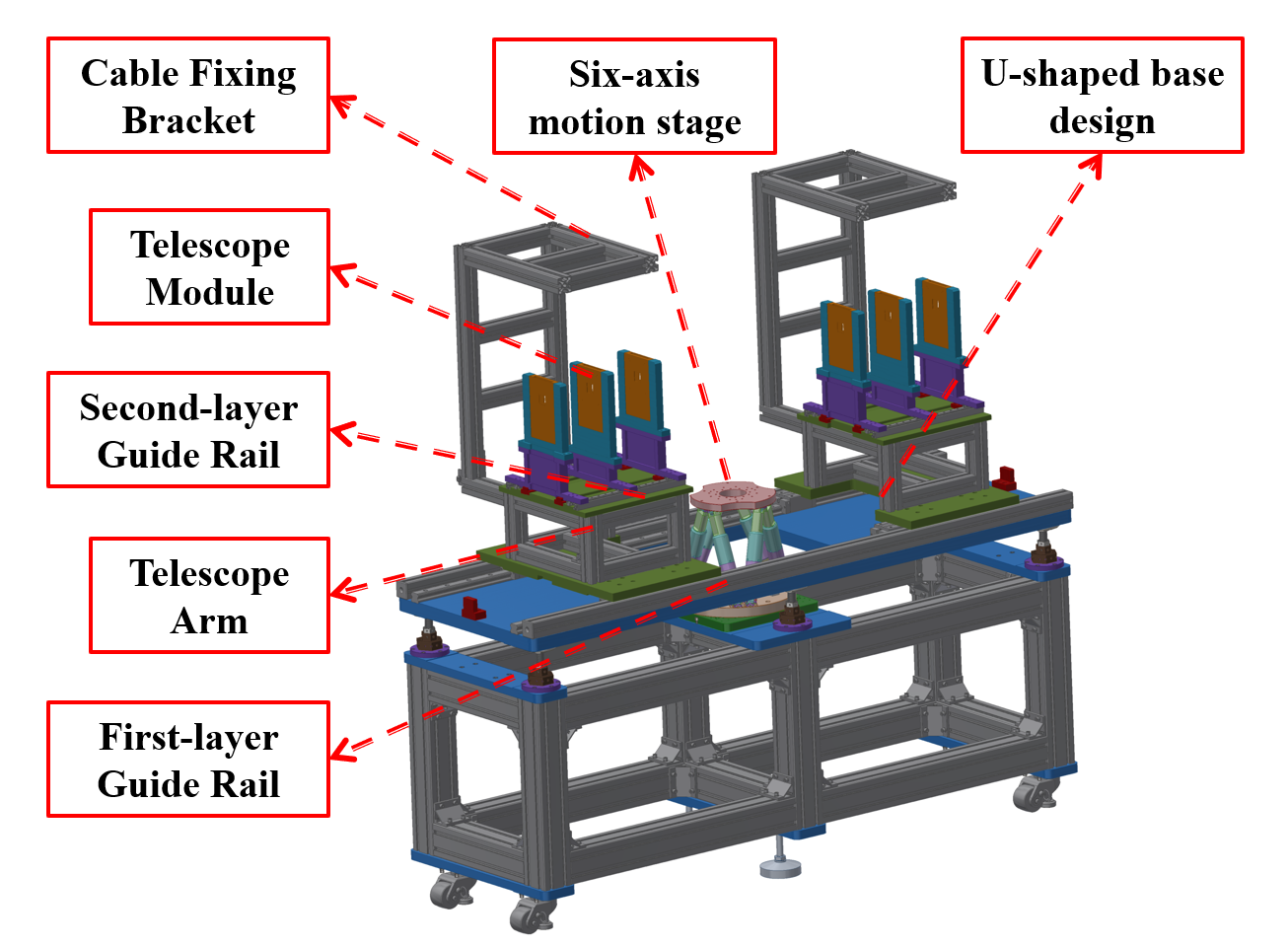}    
    \caption{Schematic illustration of the high-precision experimental platform for the beam telescope system.}
    \label{fig:2.3}
\end{figure}

A high-precision experimental platform has been designed to accommodate DUTs of various sizes and configurations. Figure~\ref{fig:2.3} shows the preliminary design of the platform, consisting of two telescope arms, each holding three telescope modules mounted on precision jigs. A two-layer guide-rail structure allows flexible repositioning of the modules along the beam axis. A six-axis motion stage is positioned at the center for DUT installation and enables remote control of its position and orientation. This platform provides accurate mechanical positioning for both the DUT and the telescope modules, forming the basis for subsequent track-based alignment.

\section{Preliminary tests}

\subsection{Radioactive source test}
\begin{figure}[!htp]
    \centering
 
    \includegraphics[width=0.3\textwidth]{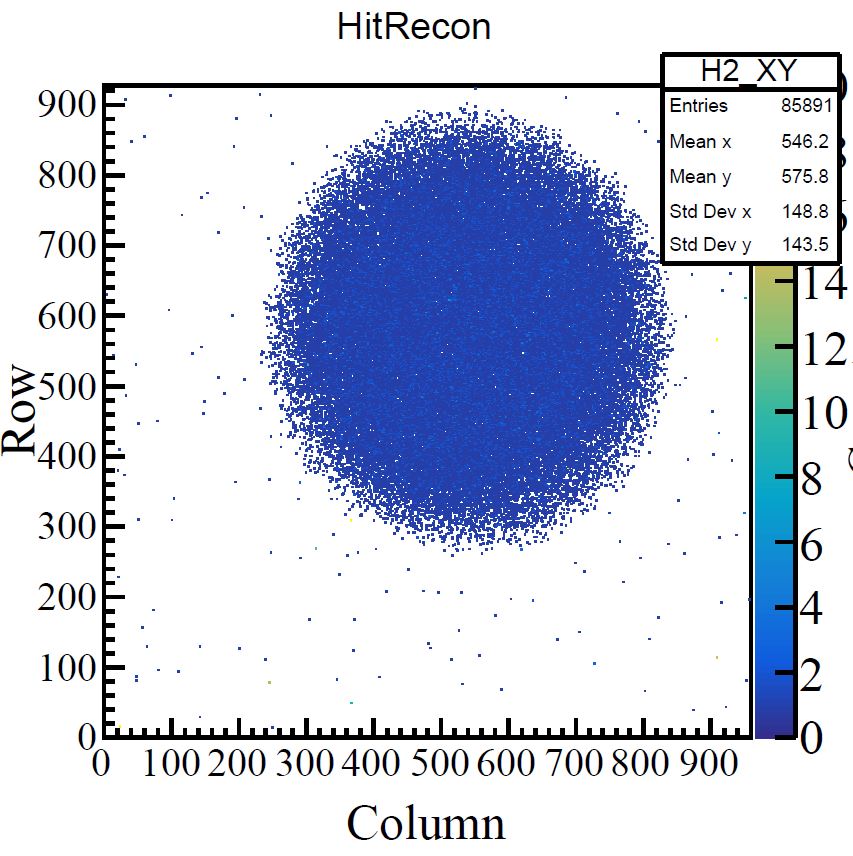} 
   
    \caption{Cluster hit-map of the telescope module exposed to a $^{55}$Fe radioactive source through a circular aperture.}
    \label{fig:3.01}
\end{figure}

The response of the telescope module was tested using a $^{55}$Fe radioactive source. Due to the charge-sharing effect, charged particles or photons incident on the telescope module can induce signals in multiple adjacent pixels, forming a cluster. The position of each cluster is calculated as the arithmetic mean of the coordinates of the fired pixels. Figure~\ref{fig:3.01} shows the cluster hit-map of the telescope module exposed to the $^{55}$Fe source through a circular collimation hole. The results demonstrate that the telescope module functions properly and that the 12.5~$\mu$m-thick black polyimide films provide sufficient light shielding during operation.

\subsection{Beam test}

To validate the design and performance of HEPTel, a preliminary beam test was carried out at the 4W1A beamline of the Beijing Synchrotron Radiation Facility (BSRF) using a 1.3 GeV electron beam. Since the TLU was not yet available, trigger signals were provided by a plastic scintillator and distributed to all modules via a fan-out board. Each shielding box was equipped with water-cooling pipes, and cooling water at 15°C was circulated to stabilize the internal temperature of the modules.

\begin{figure}[htp]
    \centering
       \includegraphics[width=0.45\textwidth]{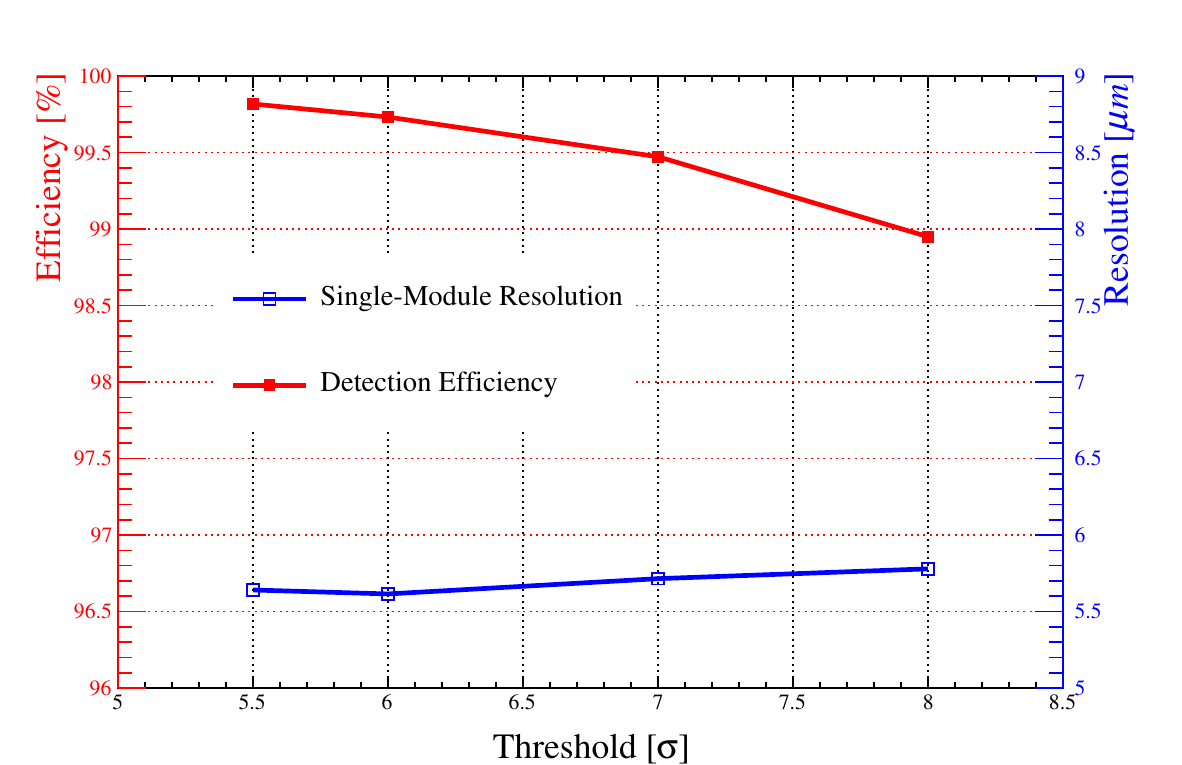} 
    \caption{The single-module resolution (blue) and the detection efficiency (red) as a function of the threshold ranging from 5.5 $\sigma$ to 8 $\sigma$.}
    \label{fig:3.4}
\end{figure}

In this test, the third telescope module served as the DUT, while the remaining five modules were used to provide reference tracks. The spacing between each module was set to 25 mm. Based on threshold scan measurements, the noise level of the telescope modules, denoted as $\sigma$, is determined to be about 18 e$^-$  (Equivalent Noise Charge). The measured residual of the DUT is approximately 6.37~$\mu$m when the threshold is set to 5.5 $\sigma$. Following the equation described in Ref.~\cite{Resolution}, this corresponds to a single-module spatial resolution of about 5.77 ~$\mu$m and a telescope resolution of about 2.70 ~$\mu$m. Figure \ref{fig:3.4} presents the single-module resolution as a function of its threshold, while the thresholds of other modules were fixed at 6 $\sigma$. The best spatial resolution was achieved with 5.5-6~$\sigma$ thresholds. The test result was slightly worse than the simulated result, mainly due to the stronger multiple Coulomb scattering of low-energy electrons and the reduced number of telescope layers used for track reconstruction. The non-monotonic dependence of the resolution on the threshold is attributed to the competition between noise suppression at low thresholds and the loss of charge sharing (smaller clusters) at high thresholds.

The single-module detection efficiency was obtained by matching reference tracks with DUT hit points. A track was counted as valid if its intersection with the DUT plane lay within 200~$\mu$m of a DUT hit. The efficiency was defined as the ratio of valid matches to the total number of reference tracks. As shown in Fig.~\ref{fig:3.4}, the efficiency stayed above 99.5\% when the threshold was set below 7~$\sigma$.

Due to the beam condition, the event rate in this test was below 100~Hz; further beam tests at higher event rates with the TLU are required to validate the telescope performance under HPES-like operating conditions.

\section{Conclusion}

Based on the requirements of HPES, a MAPS-based pixel beam telescope, HEPTel, has been designed, and its performance has been validated through simulations and preliminary beam tests. Each telescope module adopts a low material budget design of about 0.061\%~$X_0$, mitigating the effect of multiple Coulomb scattering. Simulation results indicate that, for 1.6~GeV protons, a telescope resolution of about 1.83~$\mu$m can be achieved.

A preliminary beam test using a 1.3~GeV electron beam was conducted to verify the system operation and evaluate the telescope performance. The overall telescope resolution reached about 2.70~$\mu$m, and the single-module detection efficiency exceeded 99.5\%, demonstrating the functionality of HEPTel.

Further work will focus on improving system integration, enhancing track-finding performance at high event rates, and implementing full synchronization between the telescope electronics and the TLU. 

\section{Acknowledgements}
The experiment was carried out at 4W1A beamline of Beijing Synchrotron Radiation Facility (BSRF). This work is intended for the High-Energy Proton Experimental Station of the China Spallation Neutron Source Phase-II project. This study was supported by the National Natural Science Foundation of China (No.11875274 and No.U1232202).

\end{document}